\documentclass{article}
\usepackage{amsfonts}
\usepackage{amssymb}
\usepackage{amsmath}
\usepackage{amsthm}
\usepackage[english]{babel}
\begin{document}

\title{\bf Linear perturbations in vector inflation\\ and stability issues}

\author{{\bf Alexey Golovnev}\\
{\small {\it Arnold Sommerfeld Center for Theoretical Physics, Department f\"{u}r Physik,}}\\
{\small \it Ludwig Maximilians Universit\"{a}t, Theresienstr. 37, D-80333, Munich, Germany}\\
{\small Alexey.Golovnev@physik.uni-muenchen.de}}
\date{}

\maketitle

\begin{abstract}

We continue the analysis of perturbations in vector inflation. The dominant theme of this paper is the 
long wavelength limit of perturbations in small fields inflation and the controversial issue of its linear
stability. We explain the nature of longitudinal modes, describe how they evolve, and show that they
are not as harmful as it could seem at the first glance. On the other hand, the gravitational waves instability
in large fields models is shown explicitly. It strongly limits a potential applicability of the recently
proposed $\delta N$-type approach to vector inflationary perturbations. Finally, we expose a problem of an extra (gravitational) degree
of freedom which appears whenever the vector fields are non-minimally coupled to gravity.

\end{abstract}

\vspace{1cm}

\section{Introduction}

Recently it became evident that higher spin fields can source inflationary expansion of the Universe \cite{GMV,Germani1,KMota}.
At the level of background FRW dynamics, higher spin models go almost identically to the scalar inflaton case.
But they also provide a possible account for a large scale anisotropy \cite{GMV} which is currently of a 
great phenomenological importance. It motivated many independent researchers to study cosmological perturbations in higher spin
inflationary scenarios \cite{GMV2,GV1,Lyth,Germani2}. However, the problem appears to be very complicated due to non-trivial
couplings of different types of perturbations (scalar, vector, tensor) to each other even at the linear order \cite{GV1},
with a possible exception of a 3-form inflaton field \cite{Germani1,Nunes1,Nunes2}
 being dual\footnote{Actually, the duality is flawed by higher time derivatives in the
dual action. These higher derivative effects seem 
to be intrinsically non-linear and do not spoil the linear analysis \cite{Germani2}. Probably,
they can be consistently eliminated in perturbation theory \cite{woodard}. But it is also quite possible that this problem
reveals an actual extra degree of freedom which is discussed for vector fields in Section 5.}
to some peculiar scalar \cite{Germani1,Germani2}.

In this paper we focus on vector inflation with the action \cite{GMV,GMV2,GV1}
\begin{equation}
\label{action}
S =\int\sqrt{-g}\left[-\frac{R}{2}\left(1+\sum_{n=1}^{N}\frac{1}{6}I_{(n)}\right)-
\frac{1}{4}\sum_{n=1}^{N}F_{\mu\nu}^{(n)}F_{(n)}^{\mu\nu}-\sum_{n=1}^{N}V\left(I_{(n)}\right)\right]dx^{4}
\end{equation}
where $I_{(n)}\equiv{-A}_{\mu}^{(n)}A_{(n)}^{\mu}$ and $F_{\mu\nu}^{(n)}\equiv\nabla_{\mu}A_{\nu}^{(n)}-\nabla_{\nu}A_{\mu}^{(n)}$;
the vector fields are supposed to be randomly oriented so that the background metric is approximately isotropic
with linear perturbations given by
\begin{equation}
\label{metric}
ds^{2}=\left(1+2\phi\right)dt^{2}+2a(t){\cal V}_{i}dt dx^{i}-a^2(t)\left(\left(1-2\psi\right)\delta_{ij}-h_{ij}\right)dx^{i}dx^{j},
\end{equation}
${\cal V}_{,i}^{i}\equiv 0$, $h_{i}^{i}\equiv 0,$ $h_{j,i}^{i}\equiv 0$.
The mass-term inflation corresponds to $V=-\frac{m^2}{2}A_{\mu}A^{\mu}=
\frac{m^2}{2}I$. For every single inflaton field the energy-momentum tensor takes the form
\begin{multline}
\label{EMT}
T_{\beta }^{\alpha }=\frac{1}{4}F^{\gamma \delta }F_{\gamma \delta
}\delta _{\beta }^{\alpha }-F^{\alpha \gamma }F_{\beta \gamma }+\left(2V_{,I}+
\frac{R}{6}\right) A^{\alpha }A_{\beta }+V(I)\delta^{\alpha}_{\beta}\\
+\frac{1}{6}\left( R_{\beta }^{\alpha }-\frac{1}{2}\delta _{\beta
}^{\alpha }R\right) A^{\gamma }A_{\gamma }+\frac{1}{6}\left( \delta _{\beta
}^{\alpha }\square -\nabla ^{\alpha }\nabla _{\beta }\right) A^{\gamma
}A_{\gamma }  
\end{multline}
which is by itself quite capable of giving the general feeling for why the perturbation theory is so messy. A crucial
simplification can be achieved assuming that inflation is driven by small vector fields, $NB^2\ll 1$, $B\equiv\frac{A}{a(t)}$. Anyway, 
this assumption is almost unavoidable if one wants to ensure stability of gravitational waves \cite{GMV2}. 
In Section 2 we accept it and consider the perturbations for very small values of inflaton fields.

After that we proceed with elucidating some tricky aspects of vector inflation. Namely, in Section 3 we give a thorough
analysis of longitudinal modes and associated stability problems \cite{Peloso1,Peloso2}, and in Section 4 we show the
gravitational instability of large fields vector inflation explicitly, for it has recently been doubted in \cite{Lyth}. After that,
in Section 5 we report a new problem of vector inflation concerning the number of degrees of freedom. And in
Section 6 we conclude.

\section{Linear perturbations in small fields inflation}

The full (and horrible) set of linear perturbation equations can be found in \cite{GV1}. Fortunately,
we need only a few rudiments of the general perturbation analysis. Every term in (\ref{EMT}) can be
varied easily, although the whole expression becomes very bulky. It is clear that scalar, vector
and tensor perturbations mix with each other because we can contract a background vector $B_i$ with
a perturbation. For example, using the metric (\ref{metric}) we have $\delta B^2=2B_i\delta B_i+
2B^2\psi+h_{ij}B_iB_j$. And in another linear relation, $\delta A^0=\delta A_0+{\cal V}_{i}B_i$,
the quantity ${\cal V}_{i}B_i$ is also a scalar. From $A=aB$ and $H\equiv\frac{\dot{a}}{a}$
it follows that the leading contributions
of most of the terms in (\ref{EMT}) are proportional to $H^2B^2$, with indices contracted or not.
However, at the background level the largest terms cancel each other (with $\sim\frac{1}{\sqrt{N}}$ accuracy
for $N$ random fields and exactly for the fine-tuned case of a triad), and the background dynamics
coincides with that of scalar N-flation \cite{GMV}. As for the fluctuations, one has of course to take
$H^2B\delta B$ terms into account, and we will see below that they are important.

While analysing the possible perturbations in (\ref{EMT}) step by step, it is rather tempting to
conclude that the curvature perturbations in vector inflation with a large number of fields are ridiculously small.
For example, one could argue that any terms in $T_{00}$ of the form of
 $H^2\sum A_j\delta A^j$ are statistically suppressed 
as $N\to \infty$ because the fluctuations $\overrightarrow{\delta A}$ can have 
arbitrary directions. However, it is not a reliable argument. Indeed, we definitely want to keep the Hubble
constant intact in the course of the limiting procedure (or at least the Hubble rate should not diverge). In the mass-term
inflation it means that $A\propto 1/\sqrt{N}$, and despite the $1/\sqrt{N}$ statistical suppression the length
fluctuation term $H^2\sum A_j\delta A^j$ has no scaling with $N$. It shows that one has to look for some other approximations.

Note that this type of naive argument, if it were only correct, would also suppress the perturbations in scalar N-flation
in contradiction to the general statement of \cite{LMS}. And the reason for which it does not actually happen is precisely
the same as for vectors. Namely, for $V=\frac{m^2}{2}\sum\phi_i^2$ we get $V\propto N\frac{m^2}{2}\phi^2$
and $\delta V\propto \sqrt{N}m^2\phi\langle\delta\phi\rangle$ where $\langle\delta\phi\rangle$ is a typical
magnitude (variance) of fluctuations. And then for the relative magnitude of perturbations we have
$\frac{\delta V}{V}\propto\frac{1}{\sqrt{N}}\left\langle\frac{\delta\phi}{\phi}\right\rangle$.
There is the $\sqrt{N}$ factor in denominator. However, recall that $H^2\propto m^2N\phi^2$ and hence
$\phi\propto\frac{H}{m\sqrt{N}}$. Let's parametrize the wavelengths $\lambda$ by $\kappa\equiv\left(\lambda H\right)^{-1}$ so that
$\kappa=1$ at the horizon scale. Then we have $\delta\phi_{\kappa}\propto\kappa H\propto\kappa\sqrt{N}m\phi$
and finally $\frac{\delta V_{\kappa}}{V}\propto\kappa m$. The magnitude of perturbations depends solely
on the inflaton mass. 

In vector inflation all types of perturbations are mixed, and it is particularly interesting to evaluate the effect of mixing
with gravitational waves due to $H^2\sum B_i \delta B_j$ contributions to the linear fluctuations of 
the stress tensor (\ref{EMT}). If we assume the fluctuations $\overrightarrow{\delta B}$ are random
(e.g. their directions are not correlated with the background direction of the field) then
this term has the usual $\frac{1}{\sqrt{N}}$ suppression. So that the effect should be of order $\sqrt{N}H^2B\langle\delta B\rangle$.
For the mass-term inflation we get $\delta B_{\kappa}\propto\kappa H\propto \kappa mB\sqrt{N}$, and the relative weight
of this perturbation $\sqrt{N}\langle B\delta B\rangle\propto \kappa mB^2 N$ is 
huge\footnote{This can raise some doubts about the instability of gravitational waves in such inflation
as it was deduced neglecting these terms \cite{Germani2}. Intuitively it's hard to believe that some external force
can neutralize the effect of the large tachyonic mass. And in fact, this intuition works well, see Section 4.} 
when we start inflation at $B\sim N^{-1/4}$. Some perturbations are completely out of control at the onset of large
fields inflation. It is understandable because the whole story of the background dynamics has emerged from statistical cancellation
of the leading terms $\propto H^2B^2$ in the energy-momentum tensor, and inflation is initiated when anisotropic corrections are of order one
(and actually badly unstable, see Section 4).
In new inflation the quantity $\sqrt{N}B\delta B\propto\kappa H\sqrt{N}B$ can be made relatively small.
This term serves as a source for gravity waves provided by inflatons fluctuations. One can control its magnitude
 by varying the form of potential and the number of fields in order to produce a desired amount
of tensor perturbations which is usualy a problem in new inflation (see also \cite{Vikman}).

It is very natural that, under suitable conditions in the small fields limit, the gravity waves disentangle from the other modes. After all, the linear mixing of
different perturbations occurs due to the presence of (a random set of) preferred directions. 
But the vectors which
represent the preferred directions become very small and not very important, therefore the mixing of modes is notably
weak in this limit \cite{GV1}. Hence we can approach the dynamics by considering background values of the
inflatons as small quantities. (Although they should be larger than the perturbations, of course.) For the equations
of motion of the vector fields it means that in the first approximation we are to consider the fixed background geometry, i.e. neglect the
gravitational backreaction. Indeed, all other terms in the first-order equations of motion contain
variations of metric multiplied by background vector fields, see also \cite{GV1}. 
In this limit we substitute the non-perturbed FRW metric
(with a fixed but otherwise arbitrary time-dependence of the scale factor)
in the action (\ref{action}) and get for the vector fields \cite{GMV}:
\begin{equation}
\label{1}
-\frac{1}{a^{2}}\Delta A_{0}+\left(2V_{,I}+\frac{R}{6}\right) A_{0}+\frac{1}{a}
\left(\partial _{i}\dot{B}_{i}+H\partial_i B_i\right)=0,  
\end{equation}
\begin{equation}
\label{2}
\ddot{B}_{i}+3H\dot{B}_{i}+
2V_{,I}B_{i}-\frac{1}{a^{2}}\Delta B_{i}-\frac{1}{a}\left(\partial _{i}\dot{A}_{0}+H\partial _{i}{A}_{0}\right)+
\frac{1}{a^{2}}\partial _{i}\left( \partial _{j}B_{j}\right)=0. 
\end{equation}
 And there is also a consistency condition
$$\bigtriangledown_{\mu}\left(2V_{,I}+\frac{R}{6}\right)A^{\mu}=0.$$
Note that for $H=const$ one gets a de Sitter background which amounts to setting the slow-roll
parameters to zero.

From now on we would be interested in only the leading behaviour of the vector fields and therefore assume the geometry
to be a pure de Sitter (so that the scalar curvature is constant in time) and neglect the variation of effective mass (it would
contain $V_{,II}$ and extra powers of $A$). Then we have $\bigtriangledown_{\mu}A^{\mu}=0$,
or more explicitly
\begin{equation}
\label{consistency}
\dot{A}_0+3HA_0-\frac{1}{a}\partial_i B_i=0. 
\end{equation}
It allows to simplify the equation (\ref{2}):
\begin{equation}
\label{2prime}
\ddot{B}_{i}+3H\dot{B}_{i}+
2V_{,I}B_{i}-\frac{1}{a^{2}}\Delta B_{i}+\frac{2H}{a}\partial _{i}{A}_{0}=0, 
\end{equation}
and $2V_{,I}$ can be substituted by a constant mass. We see that in the long wavelength limit all components\footnote{See
the next Section for peculiar properties of the longitudinal component which are missed in this limit.}
of $B_i$ evolve exactly like scalar fileds with the mass $m^2=2V_{,I}$. The dominant mode slowly rolls, while the other one fastly
decays. And the condition (\ref{consistency}) (or, even better, the constraint equation (\ref{1})) shows that
$A_0$ decays exponentially in physical time as it was correctly stated in \cite{GV1}. However, it is not true for the
spatial longitudinal mode itself, as it behaves identically to the transverse ones when $\lambda\to\infty$. It is just its contribution
to the constraint equation (\ref{consistency}) what goes to zero. And actually, it
goes to zero as $\frac{1}{a}$ (this dependence represents the stretching of waves) so that $A_0\propto\frac{1}{a}$ also 
(more precisely, a little bit different from that due to the slow motion of the field $B$).
Moreover, the mode which could be erroneously deduced from (\ref{consistency}) naively setting the spatial derivatives to zero,
namely $A_0\propto\frac{1}{a^3}$, is absolutely fake. The consistency condition is necessary but not sufficient 
for a vector field
to satisfy the whole system of equations of motion. From (\ref{1}) we see that if there is no longitudinal mode, then $A_0=0$ exactly.
The superhorizon analysis is always notably subtle in that it is not reliable to neglect the spatial derivatives
for evaluation of subleading quantities.

For the Einstein equations we can use the same philosophy and drop the terms which contain two powers of $B$, like $B^2\psi$;
then, for example, $\delta (B^2)$ would be given just by $2B\delta B$.
If we could also neglect the gravitational waves and consider the anisotropy as only a small correction, then the number of e-folds
in any patch of the Universe would be well-defined and depend only on variations of $B_i$ 
(recall that $A_0$ decays). In this limit and under these
assumptions one can use the $\delta N$-approach of \cite{Lyth}. However, if the anisotropies grow as fast as for the mass-term
inflation then it would be a tricky business even to speak about the number of e-folds. In the limit of small fields
the accuracy of $\delta N$-approach is determined by the strength of mixing with tensor modes. And as the latter can not be much greater 
than the tensor-to-scalar ratio (for not to produce too much of gravity waves), the $\delta N$ formalism
is applicable to at least a few percent accuracy for viable models, and we can be sure in the nearly flat primodial spectrum
of curvature perturbations if the gravitational waves are tame. Moreover, $T_{00}$ (see \cite{GV1}) depends then only on the length fluctuations
$B_j\delta B_j$ (and their time derivatives) which explains why the results of \cite{Lyth} are identical to that for
scalar N-flation. All the relevant variations in the slow-roll regime come from changing the scalar argument $A^2$
of the potential. In general, precision of these results can be roughly estimated as above,
and beyond this accuracy the perturbations differ from that of scalar inflation, although the detailed predictions are model dependent
and presumably require numerical methods.

Finally, we should mention the vector perturbations of the metric.
They usually decay as fast as $\frac{1}{a^2}$ (see for example \cite{Mukhanov}) obeying the conservation law of
angular momentum. The
vector inflation is no exception, although the decay could be a bit slower due to vector perturbations
of the energy-momentum tensor.  In absence of vectorial stress tensor perturbations the spatial part of Einstein equations
would give the usual decay. But the right hand side of it contains in vector inflation a source for ${\cal V}$.
The source represents the vortical excitations of vector fields which carry non-zero angular momentum. Therefore,
the vector inflatons can produce some vorticity. But apparently, they can not make any kind of vortical instability
at least in small fields models
because, neglecting the gravitational backreaction, the inflatons evolve in FRW space-time and by themselves obey
the angular momentum conservation, so that the source for ${\cal V}$ also decays during expansion in much the same
way as a rotating platform slows down if some mass moves on it in the radial direction. The backreaction
would allow the vector fields to gain some vorticity by transfering the opposite angular momentum to the
gravitational field as a recoil. Therefore we expect that for vector inflation the vector perturbations decay can be
a bit slower than the usual $\frac{1}{a^2}$-law, but not much slower for small fields models. It is hard to make
more precise statements on ${\cal V}$ due to the aforementioned problems of working with subleading quantities
in the long wavelength limit.

\section{The problem of longitudinal components}

At a finite wavelength, the equation (\ref{1}) and the scalar part of equation (\ref{2prime}) describe 
the evolution of temporal and longitudinal
components of vector fields. Note that (\ref{1}) is a constraint equation, and therefore
we consider only one degree of freedom per every vector field
(see, however, Section 5) which
is actually  suspected to be badly unstable for tachyonic masses \cite{Peloso1,Peloso2}. It was claimed to
be a ghost which would make any attempt to pursue a quantum field theory nonsensical, at least
in the usual perturbative formulation.
In this Section we
discuss how to live with it.

\subsection{Vector fields in flat space-time}

To set up the stage, we first describe a tachyonic ($m^2<0$) vector field in Minkowski space. We set
$R=0$, $H=0$, $B=A$ and $a=1$ in the equation of motion (\ref{2prime}) 
and get $\left(\square+m^2\right)\overrightarrow{A}=0$
for the spatial components of the vector field. 
It is just the tachyonic Klein-Gordon equation with no apparent problems. The
problem comes from equation (\ref{1}), $\left(-\bigtriangleup+m^2\right)A_0+\partial_i\dot{A}_i=0$. Fourier
decomposition in spatial modes $\propto e^{ik_i x_i}$ shows that the temporal component
\begin{equation}
\label{0}
A_0=-\frac{ik_i\dot{A}_i}{k^2+m^2}
\end{equation}
diverges at $k^2=-m^2>0$. At this wavelength the longitudinal mode is not permitted
(but $A_0$ is arbitrary) and in the neighbourhood of
$k^2=-m^2$ it involves very large values of $A_0$. As long as the tachyonic vector field is free, one can quantize
just three independent fields $A_i$ and forget about unphysical variable $A_0$ but this approach is not suitable
for turning on Lorentz invariant interactions.

In principle, the above result looks like a kind of resonant behaviour which one could try to kill by a suitable non-linearity. For example, 
it is tempting to construct a potential which would give a normal mass at large values of $A_0$. Unfortunately, given the
Lorentz invariance, the best we can do is to make it so for large positive $A_{\mu}A^{\mu}$. Then the differential operator
$-\bigtriangleup+2V_{,I}$ would still have zero modes at the length scale of $\left|m_{eff}\right|^{-1}$ 
for negative values of $A^2$ where
the field is tachyonic. The temporal component, given by (\ref{0}), would have to be very large around these
modes which is compatible with negative $A^2$ and tachyonic mass if the spatial length $\left|\overrightarrow A\right|$ 
is even 
larger\footnote{Of course, $A_0$ is by itself proportional to the length of longitudinal mode
but (in dimensions greater than $1+1$)
we can always play with the lengths of transverse components to make $A^2$ negative.}. 
Thus, the best we can do is to shift
the problem to the region of very large fields. It is not possible to completely exorcise the instability  with
modified potential, at least without abandoning Lorentz invariance or introducing new degrees of freedom.
The Hamiltonian analysis also confirms this result.

\subsection{Hamiltonian analysis}

As usual, the conjugate momenta are defined as $\pi_{\mu}\equiv\frac{\partial{\cal L}}{\partial \dot{A}_{\mu}}$
and given by $\pi_i=F_{0i}$ and $\pi_o=0$ (primary constraint). For the mass-term potential the Hamiltonian density
reads
$${\cal H}=\frac{1}{2}\pi_i^2+(\partial_i A_0)\pi_i+\frac{1}{4}F_{ij}F_{ij}+\frac{m^2}{2}\left(A_i^2-A_0^2\right).$$
Commutation of the primary constraint $\pi_o$ with the Hamiltonian gives the secondary constraint\footnote{As one
can easily check, it coincides exactly with (\ref{1}).} $\partial_i\pi_i+m^2A_0=0$ which allows for elimination of
unphysical variable $A_0$:
$${\cal H}=\frac{1}{2}\pi_i^2+\frac{1}{2m^2}\left(\partial_i \pi_i\right)^2+\frac{1}{4}F_{ij}F_{ij}+\frac{m^2}{2}A_i^2.$$
If $m^2>0$ then all the terms are positive, while for tachyonic masses the second and the fourth terms are negative. The latter
is just the ordinary tachyonic potential but the former contains momenta and therefore looks like a ghost.
However, the secondary constraint shows that for the physical states it equals just to $\frac{m^2}{2}A_0^2$:
$${\cal H}=\frac{1}{2}\pi_i^2+\frac{1}{4}F_{ij}F_{ij}+\frac{m^2}{2}\left(A_i^2+A_0^2\right).$$
There is no problem in the infrared because $A_0\to 0$ if $k\to 0$, and no problem in the ultraviolet
because $A_0\propto\frac{\dot{A}_i}{k}\propto{\cal O}(1)$ when $k\to\infty$ so that another term
$\frac{1}{4}F_{ij}F_{ij}\propto k^2A_i^2$ wins the game.
It is not a genuine ghost as it presents a problem only in a limited range of $k^2$ near $-m^2$ in a sharp contrast
with real ghosts which go worse and worse in the ultraviolet.

For a general potential the secondary constraint is $\partial_i\pi_i+2V_{,I}A_0=0$, and in terms of physical variables
the Hamiltonian becomes quite complicated since we have to use $A_0=-\frac{\partial_i\pi_i}{2V_{,I}}$ in the
argument of $V$. However, we can go the opposite way again and exclude $\partial_i\pi_i$:
$${\cal H}=\frac{1}{2}\pi_i^2+\frac{1}{4}F_{ij}F_{ij}+\frac{A_0^2}{2V_{,I}}+V(I).$$
Now one can easily convince himself that it would be rather difficult to make this Hamiltonian positive definite
because at some point we need to pass from $V_{,I}<0$ to $V_{,I}>0$.

So, we do not understand the theory completely. But it is not a fatal problem since we do not necessarily have an infinite
phase space volume in the ultraviolet for the large $A_0$ instability. And the crucial property of inflationary space-times
is that the wavelengths are changing with time and every mode dwells at a singular point $k^2=-m^2$ only for
a single instant of time. Now we proceed to show that it solves the instability problem at least at the level
of classical equations of motion\footnote{In \cite{Lyth} it is argued that even the quantum theory makes a good sense
in this situation. Of course, these arguments can be taken seriously only after we have checked that we don't really
deal with a ghost.} using the test field approximation in the simplest
case of de Sitter geometry.

\subsection{Longitudinal mode in vector inflation}

For a longitudinal mode ($k_iB_i=kB$) we use the constraint (\ref{1}) to find the closed form of the equation of motion
(\ref{2prime}) in de Sitter space ($\frac{R}{6}=-2H^2$):
\begin{equation}
\label{2deSitter}
\ddot{B}+\left(3H+\frac{2H\frac{k^2}{a^2}}{\frac{k^2}{a^2}+m^2-2H^2}\right)\dot{B}+
\left(\frac{k^2}{a^2}+m^2+\frac{2H^2\frac{k^2}{a^2}}{\frac{k^2}{a^2}+m^2-2H^2}\right)B=0.
\end{equation}
At short wavelengths there is no big difference between transverse and longitudinal modes.
The problem appears when the wavelength is near the value for which $\frac{k^2}{a^2}=2H^2-m^2$. It's time
to pay for introducing the tachyonic effective mass $-2H^2$. We need to understand the properties of solutions in this
region. Let's neglect the mass $m$ of the inflaton for simplicity as it much smaller than $H$
in the slow roll regime, and pick up a wave which was well under the horizon with
$\frac{k^2}{a^2}=4H^2$ at some instant of time $t_0$. Then $\frac{k^2}{a^2}=4H^2e^{-2H(t-t_0)}$ and 
(\ref{2deSitter}) takes the form:
$$\left(2-e^{2H(t-t_0)}\right)\ddot{B}+\left(10H-3He^{2H(t-t_0)}\right)\dot{B}+8H^2e^{-2H(t-t_0)}B=0.$$
With a new time variable $\tau=2H(t-t_0)-{\rm ln}2$ we get
\begin{equation}
\label{2reduced}
2(1-e^{\tau})\ddot{B}+\left(5-3e^{\tau}\right)\dot{B}+e^{-\tau}B=0.
\end{equation}
The critical point of crossing the singularity is at $\tau=0$. The coefficient in front of $\ddot{B}$ vanishes at this point and
all trajectories are tangent there to a one-parameter family of curves $\dot{B}=-\frac{B}{2}$. This behaviour is stable
because if $\dot{B}\neq-\frac{B}{2}$ at small $\tau<0$ then the second derivative $\ddot{B}\sim\frac{\dot{B}+\frac{B}{2}}{\tau}$
has an appropriate sign to correct the trajectory.

Our task is to find a two-parametric family of solutions for equation (\ref{2reduced}). From the previous analysis
it is clear that one possible solution contains no longitudial mode at all when $\frac{k^2}{a^2}=2H^2$.
We take $B(\tau=0)=\dot{B}(\tau=0)=0$ and construct the solution in the form of power series
$B=\tau^2+\sum\limits_{n\geq 3}C_n\tau^n$. The first term solves the equation up to ${\cal O}(\tau^2)$-terms, and
${\cal O}(\tau^3)$-corrections give $C_3=-\frac{7}{6}$ and so on. This solution is rather smooth around the
problematic point due to absence of longitudinal mode, it corresponds to $A_0=0$ at $\tau=0$.

The second solution is more interesting. We write it down as $B=1-\frac{\tau}{2}+\sum\limits_{n\geq2}D_n\tau^n$
which gives finite 
values of $A_0$ due to cancellation of two first-order zeros in (\ref{0}). The first two terms solve (\ref{2reduced}) at the
level of ${\cal O}(\tau)$, then $D_2$ is undetermined because at the level of ${\cal O}(\tau^2)$ it solves the equation by itself,
thus one can take $D_2=0$ and proceed with $D_3$. The full two-parameter family of solutions is given by
$B=\alpha-\frac{\alpha}{2}\tau+\beta\tau^2+\sum\limits_{n\geq 3}{\tilde C}_n\tau^n$ where $\alpha$ and $\beta$ are arbitrary constants while
${\tilde C}_n$-s should be determined one by one in terms of $\alpha$ and $\beta$. 

In order to understand the properties of the second solution around $\tau=0$ we note that, unlike for the first one, 
the main players for it are
$B$ and $\dot{B}$. If we neglect the $\ddot{B}$-term in (\ref{2reduced}), then the resulting equation can be solved
explicitly in elementary functions. We present the solution in terms of the physical time:
$$B=Ce^{\frac{2}{5}e^{-2H(t-t_0)}}\left(e^{-2H(t-t_0)}-\frac{3}{10}\right)^{\frac{3}{25}}.$$
One can easily check that at the moment of time $t-t_0=\frac{{\rm ln}2}{2H}$ we have $\dot{B}=-BH$ as required. It
is also straightforward to estimate the value of $A_0$ at the same time:
$$A_0\sim\frac{(\dot{B}+HB)e^{H(t-t_0)}}{(2-e^{2H(t-t_0)})H}\sim \frac{C}{20}$$
which is clearly not too large (although it is the value of a $\frac{0}{0}$ fraction), and no catastrophe happens. 
The $t\to\infty$ asymptotic is completely stable too,
the amplitude goes to a constant value (because the inflaton mass was neglected)
of the same order of magnitude as it was when we picked it up at $t=t_0$. 
(Of course, one should properly rearrange the signs for $t>t_0+\frac{1}{2H}{\rm ln}\frac{10}{3}$.)
The solutions can be evolved through the dangerous point smoothly. 

One could make the above analysis without neglecting the mass in (\ref{2prime}). It would not change the
qualitative results, but the coefficients would not be so nice. The passage through the singular 
point (with zero coefficient in front 
of $\ddot B$) gets shifted in time, but the signs of all the terms around this point 
are the same as before. Instead of the $\dot B=-H B$ behaviour at the crossing point, 
we would find $\dot B=-H\left(1+\frac{m^2}{H^2}\right) B$, and in the infinite time
limit the longitudinal mode would exhibit the slow roll
evolution $B\sim e^{-\frac{m^2}{3H}t}$. Of course, when taking into account the inflaton mass,
one should better use the actual FRW-metric instead of the pure de Sitter one, see \cite{Peloso3}.

There is no visible signature for the linear theory breakdown.
Note however that the amplitude changes from its initial value to zero, and then to 
values of opposite sign at a time scale of order of one e-fold. Of course, this is just the amplitude
of some Fourier modes around the horizon, so that it does not imply anything catastrophic, but in general
the time variations of $A^2$ in the argument of potential term can be somehow more important than usually. 
It means that non-linearity of
potential (changing of effective mass) could play some role, and therefore considerable non-Gaussianities
in primodial spectrum could be produced.

When the present article was in preparation, another work has appeared on {\it arXiv}, 
namely \cite{Peloso3}, in which the same problem is analysed. The Authors of \cite{Peloso3} 
also came to the conclusion that the linear longitudinal modes safely pass through the point 
of $k^2+M_{eff}^2=0$, at least in the case of a test field in FRW-Universe which is dubbed the zero 
vector vev in their work. They also argue that in other situations it is not the case: they claim 
that even a single vector field with non-vanishing vev in Bianchi I space-time develops a singularity 
of linear modes at this point. It is quite contrary to our general intuition which assumes that 
inclusion of the metric perturbations in (\ref{2reduced}) should not make the transition harder,
recall that it is basically governed by the sign of the coefficient in front of $\ddot B$. 
We would even expect that the transition should be softer in presence of metric perturbations since
not only different modes but also different parts of the space would pass
through this point at different instants of time. However, the 
linear instability was inferred in \cite{Peloso3} from numerical simulations. The Authors do 
not expain their numerics explicitly, but one can safely guess that they evolve 
the longitudinal mode from deep inside the horizon directly solving the equations of motion 
by standard methods which involve explicit determination of the second time derivatives of 
the physical variables via the other terms in equations. It basically amounts to solving 
our equation (\ref{2reduced}) for $\ddot B$ and using this value at each step to construct
the numerical solution. We suspect that, probably, this numerics fails to give a meaningful answer 
for a very complicated system of equations near the point at which the expression 
for the highest derivative terms is singular. On the other hand, one should also take care about
a possible anisotropic instability in the system (see the Section 4), which indeed could
change a lot.

Note also that one more problem with linear analysis was pointed out in \cite{Peloso3}. The 
linear solution diverges when $M_{eff}^2 \equiv m^2+ \frac{R}{6}=0$, approximately at the
end of inflation. We want to explain here 
the reason for that, which is actually quite simple\footnote{See also the last Hamiltonian of the Section 3.2.}. The consistency 
equation $\bigtriangledown_{\mu}\left(M_{eff}^2 A^{\mu}\right)=0$ reduces at this point 
to $\frac{dM_{eff}^2}{dt}A^0=0$. And it clearly can not be satisfied with general initial 
conditions. This is a real problem which occurs only once at the exit from inflation. 
And again, it signals the lack of fundamental understanding of the nature of the vector fields 
in the model. At this (and only this) instant of time the gauge freedom exists which makes 
the number of degrees of freedom ill-defined. The longitudinal mode becomes infinite exactly 
at the point at which it fails to be physical.

\section{Instability of large fields inflation}

In Ref. \cite{GMV2} it was shown that large fields vector inflation is badly unstable with respect
to gravitational waves, i.e. small anisotropies grow in it much too fast. It was argued in \cite{Lyth} that
this conclusion is just an artifact of the linear approximation technique in the Jordan frame, while in the Einstein frame
everything should be stable because the interaction of gravity with matter has the normal form there. However, we would like
to remind that, roughly speaking, only two thirds of the instability reported in \cite{GMV2} came from 
the $\frac{R}{6}A^2$ term in the action, while a remaining one third came from the kinetic term of the vector fields
which is conformally invariant. Of course, for this instability to develop we had to stabilize the fields
$B$ (to ensure the slow roll), but once it is done, the instability is there in both frames. In this Section we
analyse the behaviour of anisotropies in a manner which is somewhat closer to $\delta N$-formalism.

We assume that in a separate patch of the Universe the Hubble rate experienced a sudden jump in only one direction,
$H_z=H+h$.  The metric would be the axially symmetric Bianchi I type
$$ds^2=dt^2-a^2(t)(dx^2+dy^2)-b^2(t)dz^2$$
with the new components of the Einstein tensor: $G_0^0=H_a^2+2H_aH_b=3H^2+2Hh$, 
$G^x_x=G^y_y=\dot{H}_a+\dot{H}_b+H_a^2+H_b^2+H_aH_b=2\dot{H}+3H^2+\dot{h}+3Hh+h^2$ and
$G^z_z=2\dot{H}+3H^2$. If we also assume that the energy-momentum tensor remains isotropic, then subtracting
$G^z_z$ from $G^x_x$ gives the equation $\dot{h}+3Hh+h^2=0$ which shows that anisotropies
are being washed out (as they should be according to the general theorem of Wald, \cite{Wald}). If 
 $H\approx const$ then $h\propto\frac{1}{a^3}$ for small $h$. One could worry about large negative
jumps which could become stable at $h\approx -3H$, but it would already correspond to a phantom
matter with energy density $-3H^2$. And in any case, large anisotropies are not generally expected to
decay since, for example, Kasner solutions are known.

We claim that in vector inflation non-zero $h$ would render the energy-momentum tensor anisotropic too.
In the homogeneous limit and with the slow-roll assumption, the main contribution to anisotropic
part of $T_i^j$ comes from $\frac{R}{6}A_{i}A^{j}$, $-F_{0i}F^{0j}$ and $\frac{1}{6}G_i^jA^2$ terms. The
latter one renormalizes the gravitational constant, and we forget about it for a moment. But the first two terms
give important contributions to $T_x^x-T_z^z$. With a natural definition $B_z\equiv\frac{A_z}{b}$ we get for them
$$-\sum\frac{R}{6}(B_x^2-B_z^2)\approx 2NH^2\left(\langle B^2_x \rangle - \langle B^2_z \rangle\right),$$  
$$\sum(H_x^2 B_x^2-H_z^2B_z^2)\approx NH^2\left(\langle B^2_x \rangle - \langle B^2_z \rangle\right)-2NHh\frac{B^2}{3}$$
if both $h$ and $\langle B^2_x \rangle - \langle B^2_z \rangle $ are small.
And we also have to take the subleading contribution to $-F_{0i}F^{0j}$ into account, namely
$2\sum(H_x B_x\dot{B}_x-H_z B_z\dot{B}_z)$. It is easy to derive the equations of motion of vector fields:
$$\ddot{B}_x+(2H_a+H_b)\dot{B}_x+\left(m^2+\frac{R}{6}+{\dot H}_a+H_a^2+H_aH_b\right)B_x=0,$$
$$\ddot{B}_z+(2H_a+H_b)\dot{B}_z+\left(m^2+\frac{R}{6}+{\dot H}_b+2H_aH_b\right)B_z=0,$$
which show that in the slow roll regime we have
$$\dot{B}_x-\dot{B}_z\approx\frac{{\dot H}_b-{\dot H}_a+H_aH_b-H_a^2}{2H_a+H_b}B\approx\frac{\dot{h}+Hh}{3H}B$$
for a pair of identical fields in directions of $x$ and $z$ axes. And therefore
$$2\sum(H_xB_x\dot{B}_x-H_z B_z\dot{B}_z)\approx 2HB\frac{N}{3}\cdot\frac{\dot h+Hh}{3H}B\approx\frac{2NB^2}{9}(\dot h+Hh).$$
Note that at the moment of turning on the fluctuation of the Hubble rate we do not change the value of
the scale factor itself, therefore we may keep both $A$ and $B$ independent of the spatial direction of the field. 
Then the anisotropy of $T_i^j$ initially equals to $-2NHh\frac{B^2}{3}+\frac{2NB^2}{9}(\dot h+Hh)$.
Now we have to recall that the $\frac{1}{6}G_i^jA^2$ term effectively divides  the gravitational constant
by $1+\frac{NB^2}{6}$, and the resulting effective anisotropy would be $-4Hh+\frac{4}{3}(\dot{h}+Hh)$ in the large fields limit.
It has a drastic effect of making the linear equation for $h$ unstable:
$-\frac{1}{3}\dot{h}+\frac{17}{3}Hh=0$.

At the same time the anisotropy of matter distribution grows according to vector fields equations of motion:
$$\frac{d}{dt}\cdot 3NH^2\left(\langle B^2_x \rangle - \langle B^2_z \rangle\right)
\approx 3H^2\frac{N}{3}\cdot 2B\cdot \frac{\dot{h}+Hh}{3H}B\approx \frac{2HNB^2}{3}\left(\dot{h}+Hh\right),$$
and in this case tends to stabilize the fluctuation. However, if we differentiate the anisotropic part of Einstein equations
with respect to time:
$$\frac{d}{dt}\left(-\frac{1}{3}\dot{h}+\frac{17}{3}Hh\right)\approx\left(1+\frac{NB^2}{6}\right)^{-1}\cdot \frac{2HNB^2}{3}\left(\dot{h}+Hh\right),$$
we get a simple approximate equation $\ddot h-5H\dot h+12H^2h=0$ with initial condition $\dot{h}(0)=17Hh(0)$. Its solution
exhibits a fast exponential growth with oscillations.

Admittedly, the geometric effects which overturned the sign of $\dot h$ can be changed by, for example, taking anisotropic distributions
of the vector fields. But even if one can force $h$ to decay initially, then after a small period of time
(before the perturbation could finally decay)
one generically gets a considerable amount of anisotropy in the term which does 
not depend on the current value of $h$, namely $3NH^2\left(\langle B^2_x \rangle - \langle B^2_z \rangle\right)$.
This conclusion is robust. It can either prevent the Hubble rate jump from stopping at zero value and make it growing with the
opposite sign, or change the sign of $\dot h$ before the fluctuation could reach the zero, depending on the coefficients in other terms.
In any case, a small anisotropic fluctuation in the Hubble law is badly unstable for the large fields vector inflation.

\section{An extra degree of freedom}

Now we want to report on a new problem with vector inflation. A naive counting of independent propagating degrees of freedom
would give the number of $3N+2$ for $N$ vector fields and  one graviton. Equations also contain the second time derivatives
of $\psi$. But it is usually not a dynamical quantity since the temporal component of the Einstein
equations contains no second time derivatives\footnote{One notable exception is a treatment by Weinberg \cite{Weinberg} who
casts the equations into a form $R_{\mu\nu}=T_{\mu\nu}-\frac{1}{2}g_{\mu\nu}T^{\alpha}_{\alpha}$ which contains time
derivatives in the temporal component too. (Weinberg has also an overall minus sign in the right hand side due to the opposite
sign convention for the Ricci tensor.) Of course, a proper linear combination of these equations restores
the constraint.} and can be (and should be) regarded as a constraint which determines
$\psi$ completely. (It actually corresponds to Newtonian potential defined by distribution of masses
in non-relativistic limit.) However, for vector inflation it is not the case beyond the linear approximation
since $T_{00}$ has a term with $RA_0^2$ and the scalar curvature depends on $\ddot{\psi}$. On the other hand,
the field equation (\ref{1}) is also no longer
a constraint due to precisely the same reason. And it is not clear how to find those linear combinations of equations which would give
constraints in the Jordan frame.

To find out the actual number of degrees of freedom we need to perform
a conformal transformation to the Einstein frame:
$\tilde{g}_{\mu\nu}=e^{2\rho}g_{\mu\nu}$,
$\tilde{R}=e^{-2\rho}\left(R-6\square\rho-6(\nabla\rho)^{2}\right)$, $\tilde{A}_{\mu}^{(n)}={A}_{\mu}^{(n)}$ 
with the conformal weight
$$\rho=\frac{1}{2}{\rm ln}\left(1+\frac{1}{6}\sum_{n=1}^NI_{(n)}\right).$$
It transforms the original action to\footnote{In the slow roll regime the conformal factor changes slowly
with time, so that the scalar quantities $A^2$ and $F^2$ remain almost frozen in both frames. The new kinetic
term $\left(\bigtriangledown\varphi\right)^2$ plays the role of non-minimal coupling
ensuring the slow roll conditions.}
\begin{equation}
\label{conformal}
S=\int dx^{4}\sqrt{-g}\left(-\frac{R}{2}+\frac{1}{2}\nabla_{\mu}\varphi\nabla^{\mu}\varphi
-{\bf V}-\sum_n\frac{1}{4}F^{(n)}_{\mu\nu}F_{\alpha\beta}^{(n)}g^{\mu\alpha}g^{\nu\beta}\right)
\end{equation}
where
$$\varphi=\sqrt{\frac{3}{2}}\cdot{\rm ln}\left(1+\frac{1}{6}\sum_n I_{(n)}\right)$$
and the new potential is given by
$${\bf V}=\left(1+\frac{1}{6}\sum_mI_{(m)}\right)^2 \cdot\sum_n V\left(\left(1+\frac{1}{6}\sum_pI_{(p)}\right)^{-1}\cdot I_{(n)}\right),$$
so that the vector fields are intricately interacting now. This potential looks scary, and one can think of
the special case $V=\frac{m^2}{2}I$ for which it is just 
$${\bf V}=\frac{m^2}{2}\left(1+\frac{1}{6}\sum_mI_{(m)}\right)\cdot\sum_n  I_{(n)}.$$
Anyway, our discussion does not depend on a potential. It refers to any non-minimally coupled vector fields
(and should be suitable for higher forms too).

We first consider the new action (\ref{conformal}) with $N=1$ in a flat (Minkowski) space-time. Canonical momenta
are given by $\pi_0=4A_0\varphi^2_{,I}\left(A_0\dot{A}_0-A_j\dot{A}_j\right)$ and
$\pi_i=-4A_i\varphi^2_{,I}\left(A_0\dot{A}_0-A_j\dot{A}_j\right)+\dot{A}_i-\partial_i A_0$. This system of linear differential equations
for velocities has a determinant $$-\left(4\varphi^2_{,I}\right)^3A_0^2\left(A_1^2A_2^2+A_2^2A_3^2+A_3^2A_1^2\right)$$
which is not zero if $A_0\neq0$ and therefore is solvable under this assumption, and no constraints are there.
Thus, the vector field acquires a fourth degree of freedom whenever the longitudinal mode is on. The reason is easy to
understand analysing the equation of motion
\begin{equation}
\label{motion}
2A^{\nu}\frac{\partial\varphi}{\partial I}\left(\square\varphi\right)+
\nabla_{\mu}F^{\mu\nu}+2A^{\nu}\frac{\partial {\bf V}}{\partial I}=0.
\end{equation}
Its temporal component  is not a constraint due to the time derivatives in $\square\varphi$.

For an arbitrary number of fields, evaluation of the determinant is a tedious procedure. $3N+1$ degrees of freedom are guaranteed,
but a priori there could be even more of them, up to $4N$. However, we can check that there is only one extra degree of freedom
at any value of $N$ by uncovering $N-1$ constraints. It is easy to do. Suppose, we have two fields  with equations
of motion (\ref{motion}). Let's multiply the temporal component of the first equation by 
$2A^{0}_{(2)}\frac{\partial\varphi}{\partial I_{(2)}}$, the temporal component of the second equation by
$2A^{0}_{(1)}\frac{\partial\varphi}{\partial I_{(1)}}$ and subtract them from each other. Time derivatives are
 cancelled, and we get a constraint. Then we can play the same game with other fields. Clearly,
there are $N-1$ independent relations of this sort.

We see that whenever at least one of the fields has $A_0\neq 0$, an extra degree of freedom
turns on and completely decouples again
in absence of longitudinal modes. It can't be seen at all at the level of background dynamics. Although this mode should
not play any role in the linear perturbation analysis around the state with $A_0=0$, it means that at the
fundamental level the theory is not very well defined. 
The problematic degree of freedom is essentially gravitational because
it is one for all the fields and stems from non-minimal coupling term due to time derivatives in $R$. Perhaps, it could
be regulated by explicitly turning it on even in the $A_0\to 0$ limit, for example with a small $R^2$-correction
to Einstein gravity.

\section{Conclusions}

An interesting possibility of driving inflation with higher spin fields is nowadays being emergent. It appears that
this idea can be perfectly realised at the background level with almost no complications \cite{GMV,Germani1,KMota} and with a benefit of a natural large scale
anisotropy suggested by some recent observations \cite{erik}. But already at the level of linear perturbations the full set
of equations of motion \cite{GV1} becomes almost analytically untractable (with a possible exception
of 3-form inflation \cite{Germani1,Germani2,Nunes1,Nunes2}). And also some problems have been 
reported both before \cite{Peloso1,Peloso2,Peloso3,Chiba}
and now, in this paper, see Section 5. However, we have shown that although there is obviously a serious lack of understanding
the nature of vector inflatons at the fundamental level (let alone the problem of UV-completion), nevertheless
one can consistently use small fields inflation as an effective theory of the early cosmological evolution.

Very recently, a first calculation of non-Gaussianities for inflation with vector fields has
appeared, see \cite{newLyth}. It is claimed that vector fields can produce a high level
of non-Gaussianity. The analysis in \cite{newLyth} is performed in the usual $\delta N$-formalism
of \cite{Lyth} with all its potential shortcomings, but the possibility
is nevertheless very interesting. Note also that among the things which could not been taken into account in \cite{newLyth} is
the tricky evolution of longitudinal modes. We have seen above that this evolution is
very peculiar, and therefore a general intuition would imply that it also can
make a contribution to non-Gaussian features. Of course, this issue needs a further investigation.

We conclude that there are many unresolved fundamental issues about vector (and higher spin) inflation, but at the level
of effective description it is a viable candidate for the theory of inflationary epoch. Moreover, it could provide
interesting insights into quite a few problems of current phenomenological interest. A great deal of further
progress remains to be done and, from our perspective, worth to be done in the field of understanding vector inflation.

The Author is grateful to Vitaly Vanchurin, Viatcheslav Mukhanov, David Lyth, and Cristiano Germani 
for very useful discussions.
This work was supported in
part by the Cluster of Excellence
EXC 153 {}``Origin and Structure of the Universe''.

\end{document}